\documentclass[12pt]{article}
\usepackage{graphicx}

\newcommand\pubnumber{ }
\newcommand\pubdate{\today}

\def\nMSM{$\nu MSM$}
\def\GeVc2{GeV/$c^2$}
\def\mN{$m_N~$}

\def\INFNpg{Sezione INFN di Perugia, I-06123 Perugia, ITALY}
\def\support{\footnote{on behalf of the NA62 Collaboration: R. Aliberti, F. Ambrosino, R. Ammendola, B. Angelucci, A. Antonelli, G. Anzivino,R. Arcidiacono, M. Barbanera, A. Biagioni, L. Bician
% ...
}}

\def\Title#1{\begin{center} {\Large #1 } \end{center}}
\def\Author#1{\begin{center}{ \sc #1} \end{center}}
\def\Address#1{\begin{center}{ \it #1} \end{center}}

\newcommand\pubblock{\rightline{\begin{tabular}{l} \pubnumber\\
         \pubdate  \end{tabular}}}
\newenvironment{Abstract}{\begin{quotation}  }{\end{quotation}}
\newenvironment{Presented}{\begin{quotation} \begin{center} 
             PRESENTED AT\end{center}\bigskip 
      \begin{center}\begin{large}}{\end{large}\end{center} \end{quotation}}

\def\beq{\begin{equation}}
\def\eeq#1{\label{#1}\end{equation}}
\def\eeqn{\end{equation}}

\def\beqa{\begin{eqnarray}}
\def\eeqa#1{\label{#1}\end{eqnarray}}
\def\eeqan{\end{eqnarray}}

\let\bar=\overbar

\def\Dslash{\not{\hbox{\kern-4pt $D$}}}
\def\dslash{\not{\hbox{\kern-2pt $\del$}}}

\def\msb{{\bar{\ssstyle M \kern -1pt S}}}

\begin{document}
\begin{titlepage}
\pubblock

\vfill
\Title{Searches for Heavy Neutrinos at the CERN SPS}
\vfill
\Author{ Patrizia Cenci\support}
\Address{\INFNpg}
\vfill
\begin{Abstract}
Searches for heavy neutrinos can be successfully performed by fixed target experiments at the CERN SPS. 
New results obtained by the NA62 and NA48/2 kaon experiments are summarized in this paper.
The physics potential of future projects exploiting SPS protons with beam dump facilities to enlarge the sensitivity to heavy neutrinos above the kaon mass limit are also outlined.
\end{Abstract}
\vfill
\begin{Presented}
NuPhys2017, Prospects in Neutrino Physics\\
Barbican Centre, London, UK,  December 20--22, 2017
\end{Presented}
\vfill
\end{titlepage}
\def\thefootnote{\fnsymbol{footnote}}
\setcounter{footnote}{0}

\section{Introduction}
The evidence for massive neutrinos, due to the observation of neutrino oscillations,
requires to extend the Standard Model (SM) to accommodate neutral lepton masses. 

A natural extension of the SM predicts new sterile neutrinos, here called heavy neutrinos (HNs), which mix with the ordinary light neutrinos. In the Neutrino Minimal Standard Model (\nMSM) \cite{MSMn1}\cite{MSMn2} three massive right-handed singlet neutrinos, with masses comparable to those of the SM quarks and charged leptons, are introduced to explain simultaneously neutrinos oscillation, dark matter and baryon asymmetry in the Universe (BAU). The lightest HN, with mass of O(10 keV/$c^2$)), is a dark matter candidate. The other two HNs have masses of O(1 \GeVc2), produce the SM neutrino masses through the see-saw mechanisms and introduce extra CP violating phases to account for BAU.
A mixing matrix U describes the interactions between HNs and SM leptons \cite{HNK1}\cite{HNK2}. 
The mass range and the small mixing angles predicted in the \nMSM~make HNs long lived, with mean free paths of O(10 km) or longer, and production branching fractions of O($10^{-10}$) or smaller. 

The mixing of sterile and active neutrinos leads to the production of HNs in weak decays of hadrons, making them accessible in beam-beam and beam-target collisions at accelerator-based experiments. The same mixing provides their decays to SM particles \cite{HNTh}. 
Due to the smallness of couplings and the consequent long lifetimes, luminosity and acceptance limit the HN reach of present experiments.
For the same reason, the production of the lightest sterile neutrino is negligible and only the two heaviest ones can be detected in accelerator-based experiments. 
The current strongest bounds on HN couplings with SM particles have been established up to the kaon mass limit, by experiments searching for HN production in kaon and pion decays . 

New results from HN searches in charged kaon decays, recently set by the NA48/2 and NA62 experiments at the CERN SPS, will be summarized in this paper. The physics potential of future projects exploiting SPS protons with beam dump facilities to enlarge the sensitivity to HN above the kaon mass limit will also be briefly reported.

\section{HN searches at CERN SPS: kaon experiments}
Searches for HNs in the kaon sector are among the most promising for a large part of the parameter phase space. 
The mixing between HNs and SM neutrinos leads to HN production in the leptonic kaon decays $K^\pm\rightarrow l^\pm N$ $(l=e, \mu)$, where N is the new HN. 
The branching ratio (BR) of those decays is determined by the HN mass \mN and the mixing matrix parameter $|U_{l4} |^2$. This BR can be expressed in terms of the BR of the SM semileptonic decay $K^\pm\rightarrow l^\pm \nu$ ($K_{l2}$) using a kinematic factor $\rho_l$($m_N$) accounting for helicity suppression and phase space dependence on $m_N$\cite{HNK1}\cite{HNK2}:
\begin{equation} \label{eq:BRequation}
BR(K^\pm \rightarrow l^\pm N) = BR(K^\pm \rightarrow l^\pm \nu) \times \rho_l(m_N) \times |U_{l4}|^2
\end{equation} 
Limits on the mixing matrix elements  for \mN below the kaon mass can be established by searching for peaks in the missing mass spectrum of kaon leptonic decays.

Complementary studies have been performed by the NA48/2 and NA62 experiments at CERN, exploiting charged kaon decays.
NA48/2 searched for HNs looking into both production and decay mechanisms with the data collected in 2003 and 2004 \cite{NA48/2/HN}.
In NA62 the data taken in 2007 at an early stage of the experiment, and those collected later with a new experimental apparatus in 2015, have been analyzed to search for HN production in leptonic kaon decays \cite{NA62/RK/HN}\cite{NA62/2015/HN}. The analysis of 2016 and 2017 data is well advanced and further improvements of the NA62 results are expected soon.

\subsection{NA48/2 Results}
The largest world sample of charged kaon decays has been collected by the NA48/2 experiment, which took data in 2003 and 2004 with the main goal of searching for direct CP violation in 3-pion decays \cite{NA48/2/CP}. 
Simultaneous $K^+$ and $K^-$ beams with (60$\pm$3) GeV/$c$ momentum were produced by 400 GeV primary protons from the CERN SPS impinging on a beryllium target. 
The fiducial decay volume was contained in a cylindrical vacuum tank, 114 m long. 
Charged particle momenta were measured by a magnetic spectrometer composed of a dipole magnet and four drift chambers located in a vessel filled with helium gas. The spectrometer was followed by a scintillator trigger hodoscope and a liquid krypton electromagnetic calorimeter (LKr). A hadron calorimeter and muon veto system were located further downstream to distinguish muons from pions. 
Details of the experimental apparatus are available in \cite{NA48/2/Det}.

The search for production and decay of HN into a pair of opposite charge particles has been exploited by NA48/2 on events with three tracks in the final state \cite{NA48/2/HN}. 
A sample of $K^{\pm} \rightarrow \pi\mu\mu$ decays ($K_{\pi\mu\mu}$) has been analyzed looking for short-living sterile neutrinos $N_4$ produced in $K^{\pm} \rightarrow \mu^{\pm} N_4$ transitions and promptly decaying into charged pions and muons according to the process $N_4 \rightarrow \pi\mu$.  

\begin{figure}
\centering
\includegraphics[height=2.2in]{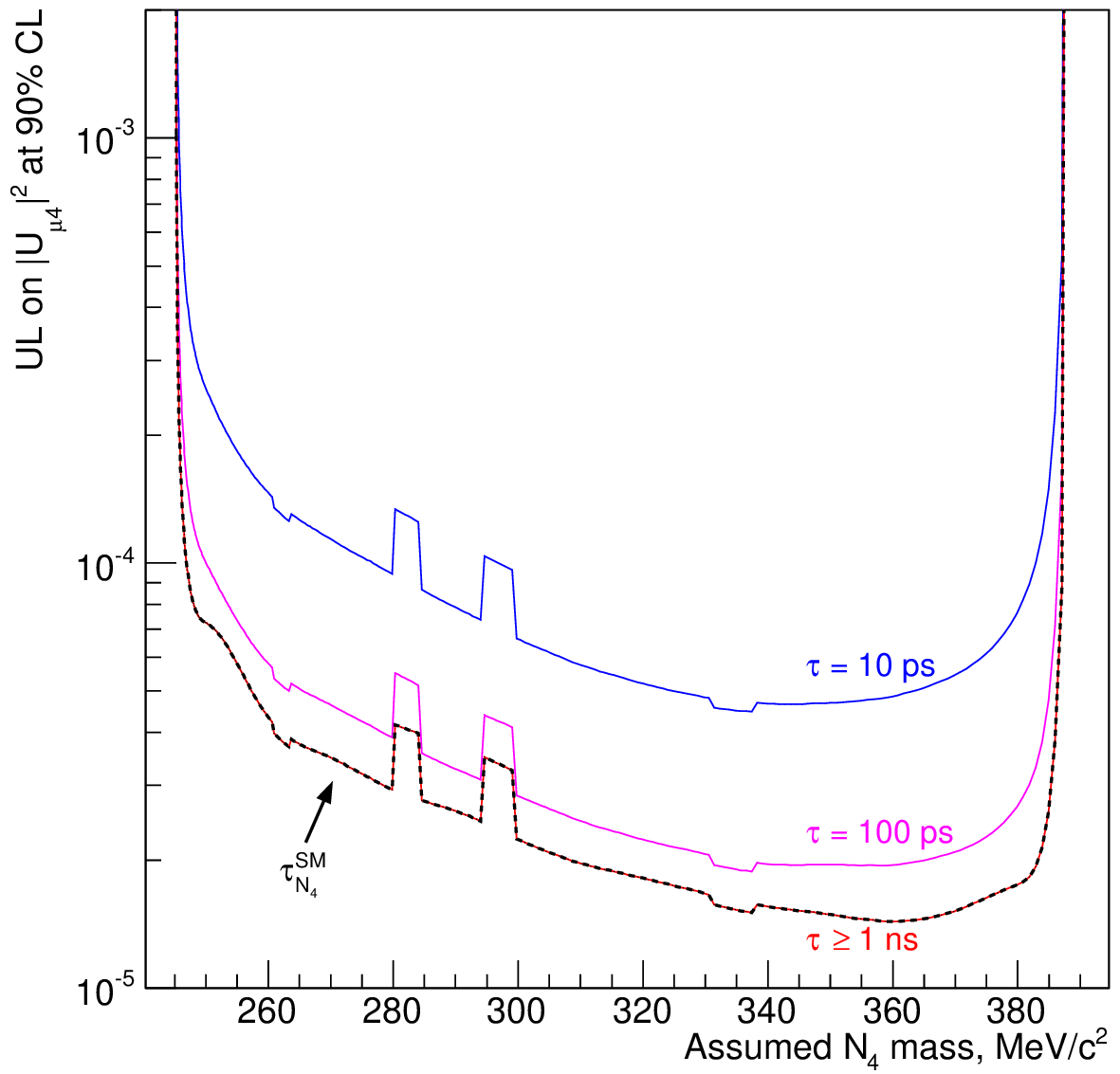}~~~~~~~
\includegraphics[height=2.2in]{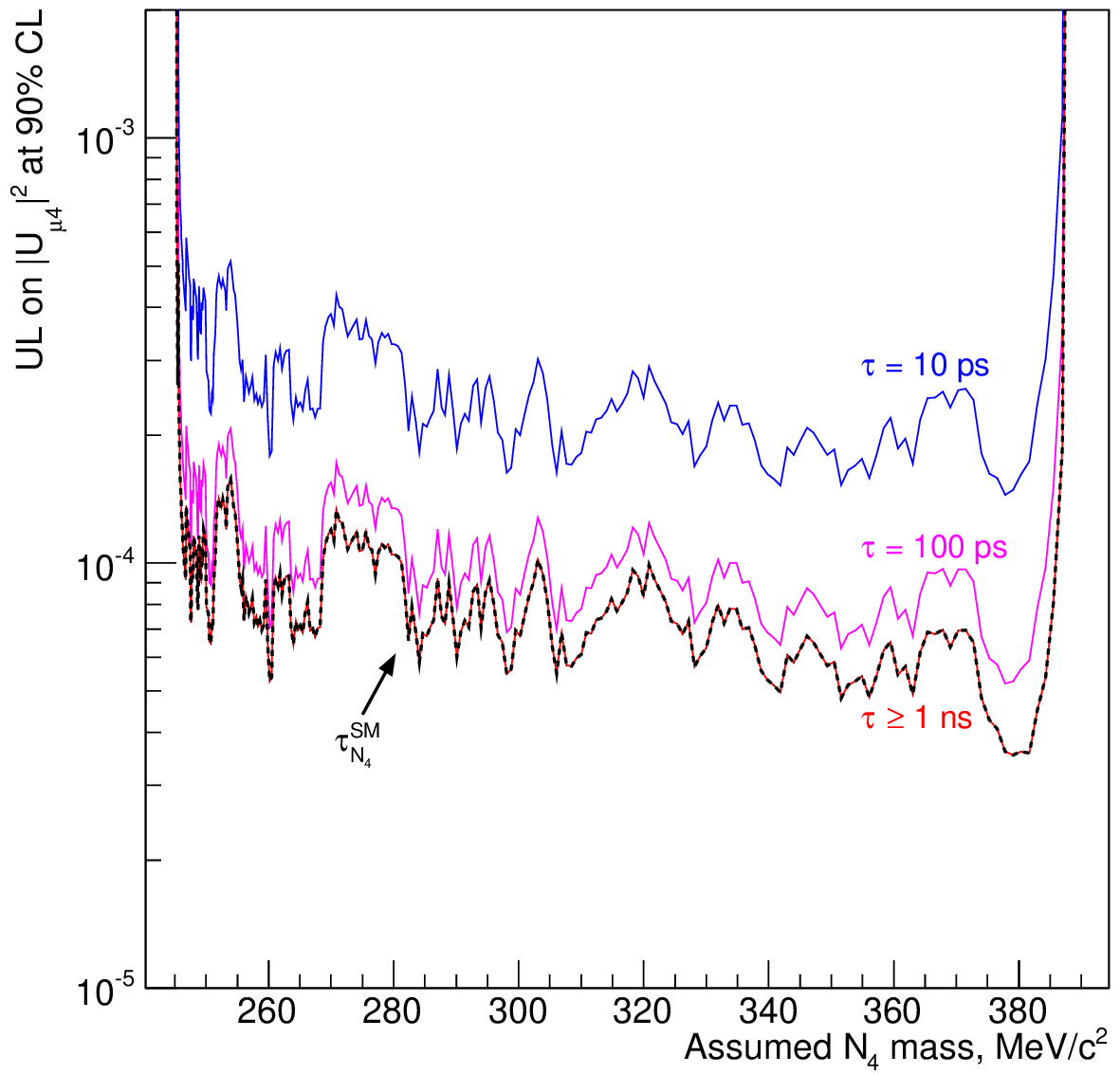}
\caption{NA48/2 upper limits ($90\% ~C.L.$) on the squared matrix element $|U_{l4}|^2$ as a function of the HN mass and lifetime $\tau$ for LNV (left) and LNC (right) $K_{\pi\mu\mu}$ decays.}
\label{fig:na48u}
\end{figure}

A peak in the invariant mass distribution $M_{\pi\mu}$ of the pion-muon system could reveal the existence of the new sterile neutrino $N_4$.
The result depends on the assumptions made on $N_4$ lifetime.
Two categories of events have been studied: the Lepton Number Violating (LNV) decay $K^{\pm} \rightarrow \pi^{\mp}\mu^{\pm}\mu^{\pm}$ and the Lepton Number Conserving (LNC) decay $K^{\pm} \rightarrow \pi^{\pm}\mu^{\pm}\mu^{\mp}$. The first one, forbidden in the SM, could be mediated by Majorana neutrino exchange. 
$K_{\pi\mu\mu}$ event rates are measured relative to the $K^{\pm} \rightarrow \pi^{\pm}\pi^{\pm}\pi^{\mp}$ ($K_{3\pi}$) normalization channel. 
The analysis is performed on a sample of $2 \times 10^{11}$ kaon decays estimated from the $K_{3\pi}$ sample.
Signal and normalization channels are collected concurrently through the same trigger logic. They have similar topologies, since muon and pion masses are close to each other, leading to first order cancellations of systematic effects due to trigger efficiency and geometrical acceptance.
The event selections require a well reconstructed 3-track vertex and share many common parts. Particle identification and kinematic criteria distinguish $K_{\pi\mu\mu}$ candidates at a later stage.
Track momenta between 5 and 55 GeV/$c$ are selected. The total momentum must have a longitudinal component compatible with the beam momentum and a negligible transverse component. Pion and muon identification is achieved by exploiting the muon veto detector information and the energy deposited in the LKr calorimeter compared with the momentum measured by the spectrometer.
The invariant mass $M_{\pi\mu\mu}$ is computed for events with the decay vertex in the fiducial volume.
The signal region is defined with respect to the nominal kaon mass $M_K$ as $| M_{\pi^{\mp}\mu^{\pm}\mu^{\pm}} - M_K | <$ 5 MeV/$c^2$ and $| M_{\pi^{\pm}\mu^{\pm}\mu^{\mp}} - M_K| <$ 8 MeV/$c^2$ for LNV and LNC events respectively. Different selections are applied due to different background contaminations in the two samples.
One event is observed in the signal region for the LNV event selection. 
A background expectation of $N_{bkg} = 1.16 \pm 0.87_{stat} \pm 0.12_{syst}$ has been estimated with Monte Carlo simulation.
The upper limit $BR(K^{\pm} \rightarrow \pi^{\mp}\mu^{\pm}\mu^{\pm}) < 8.6 \times 10^{-11}$ is set at $90\% ~C.L.$~.
A sample of 3489 $K^{\pm} \rightarrow \pi^{\pm}\mu^{\pm}\mu^{\mp}$ candidates has been found in the signal region after the LNC event selection, with an estimated background contamination of $(0.32 \pm 0.09)\%$.	

A scan of the invariant mass $M_{\pi\mu}$ is performed for both LNV and LNC decays to search for 2-body resonances over a range of mass hypotheses in steps given by half of the mass resolution $\sigma(M_{\pi\mu})$. 
The statistical analysis makes use of the Rolke-Lopez method to find 90$\%$ confidence intervals for Poisson processes with multiple Poisson background with unknown mean.
The significance of the signal does not exceed 3 standard deviations in any mass bin. 
The upper limits on the squared matrix magnitude $|U_{\mu 4}|^2$ depend on the HN mass \mN and lifetime $\tau$ assumptions and have been evaluated from those set on the products  
$BR(K^{\pm} \rightarrow \mu^{\pm} N_4) \times BR(N_4 \rightarrow \pi^\mp \mu^\pm)$ and
$BR(K^{\pm} \rightarrow \mu^{\pm} N_4) \times BR(N_4 \rightarrow \pi^\pm \mu^\mp)$.
Figure~\ref{fig:na48u} shows the upper limits on $|U_{\mu 4}|^2$ as a function of the HN mass hypothesis and the lifetime $\tau$ for, respectively, LNV and LNC transitions. 
The results are valid for unstable HNs with lifetime $\tau$ $<$100 $\mu s$.

\subsection{NA62 Results}
\subsubsection{NA62 2007 data}
NA62 is the last generation kaon experiment at the CERN SPS and aims to measure the BR of the ultra-rare $K^+ \rightarrow \pi^+ \nu \overline{\nu}$ decay with 10$\%$ accuracy. 
It has been taking data at an early stage, in 2007, to test lepton flavor universality by measuring the ratio $R_K$=$\Gamma(K_{e2})$/$\Gamma(K_{\mu2})$ of the widths of charged kaon leptonic decays \cite{NA62/RK}.
The experimental apparatus was based on the NA48/2 detector and beam line, optimized to improve the collection of decays with electrons and muons in the final state.
The beam momentum was ($74 \pm 1.4$) GeV/$c$, the intensity was reduced to exploit a highly efficient minimum bias trigger configuration with minimal accidental background.

In 2007 NA62 collected about $2 \times 10^{10} ~K^\pm$ decays.
The production of a new sterile neutrino N has been investigated in $K^+\rightarrow \mu^+ N$ transitions due to a smaller muon halo background contamination in the $K^+$ sample \cite{NA62/RK/HN}. 
Assuming $|U_{\mu4}|^2 < 10^{-4}$ and HN decays into SM particles, the HN mean free path is above 10 Km and its decay probability in the NA62 fiducial volume is negligible.
The events are selected by requiring one single track in the detector acceptance with positive charge and momentum between 10 and 65 GeV/$c$, identified as a muon coming from a vertex in the decay region. 
The expected HN signature is a peak in the distribution of the event squared missing mass defined as $m^2_{miss} = (P_K - P_\mu)^2$, where $P_\mu$ is the 4-momentum of the track reconstructed in the spectrometer assumed to be a muon, and $P_K$ is the kaon average momentum as monitored from reconstructed $K_{3 \pi}$ events.
Simulated and measured $m_{miss}$ distributions are compared to set limits on the number of observed $K^+\rightarrow \mu^+ N$ decays at different mass values. The mass signal region is 300$-$375 MeV/$c^2$. Although accessible, lower masses have not been considered since strong upper limits of $O(10^{-8})$ have already been set on $|U_{\mu4}|^2$ below 300 MeV/$c^2$ \cite{E949}. 

\begin{figure}
\centering
\includegraphics[height=2.1in]{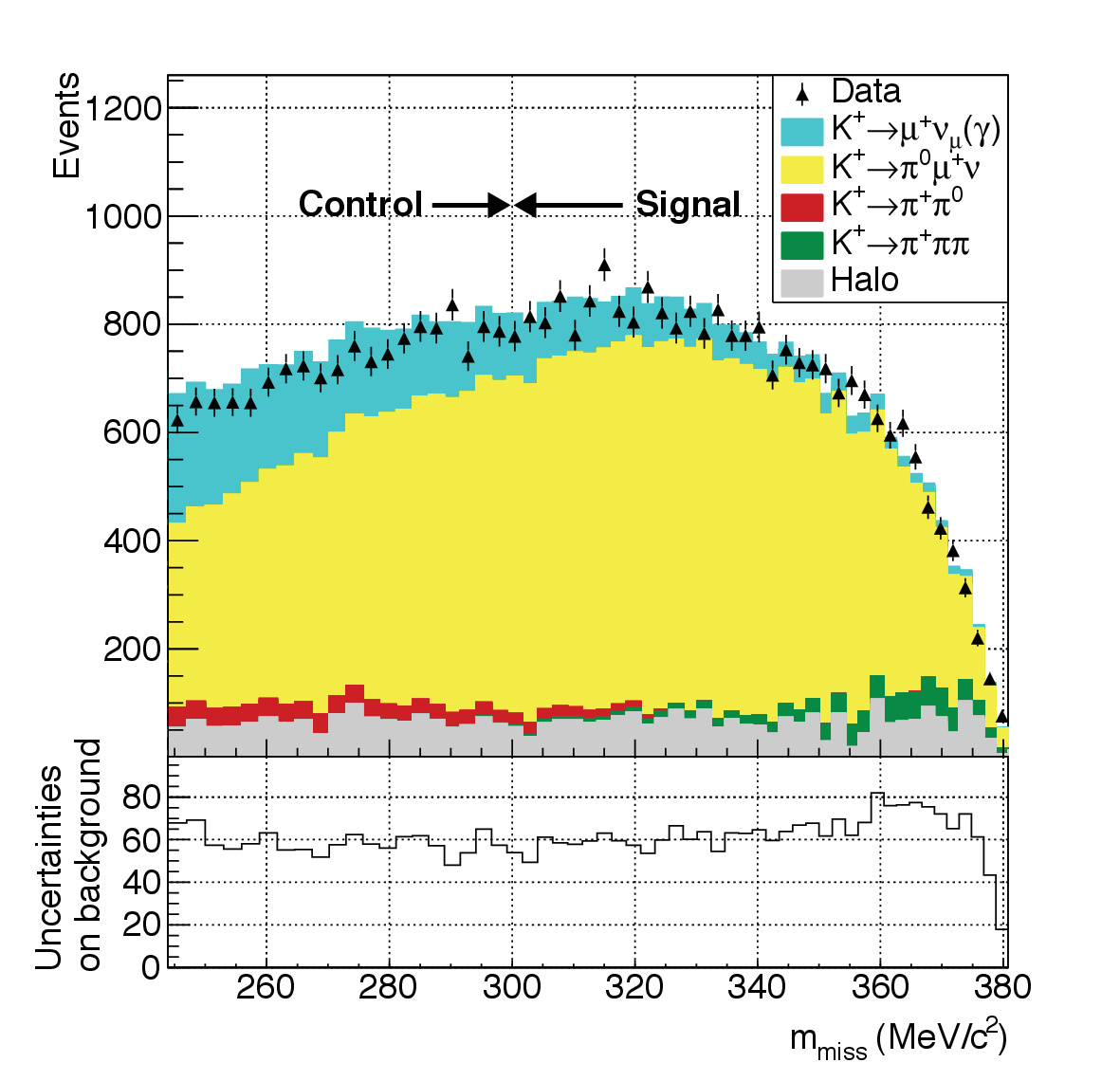}~~~~~~~~
\includegraphics[height=2.1in]{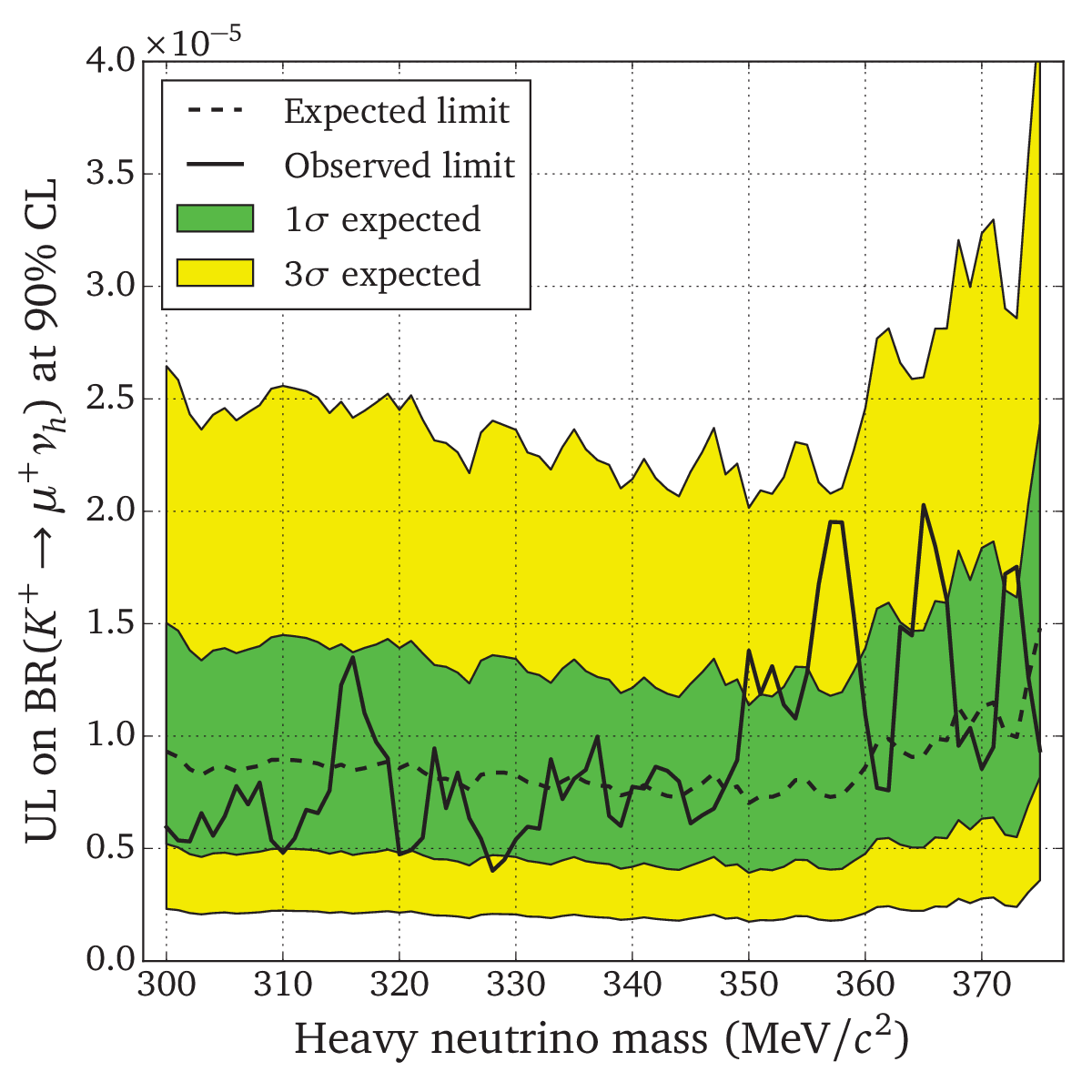}
\caption{NA62, 2007 data. Left: missing mass distributions for data, with statistical uncertainties, and for background contributions in both signal and control regions. The lower plot shows the total uncertainty on the background estimate.
Right: expected and observed upper limits ($90\%~C.L.$) on BR($K^{\pm} \rightarrow \mu^{\pm} N$) at each HN mass.
}
\label{fig:na62-2007}
\end{figure}

A data set of $N_K = (5.977 \pm 0.015) \times 10^7$ kaon decays has been analyzed.
The background is mostly produced by beam halo muons, studied with a control sample of $K^-$ decays, and by kaon decays evaluated with simulation.
Figure~\ref{fig:na62-2007} (left) shows the $m_{miss}$ distribution of the selected events and the background contributions. 
The largest relative systematic uncertainty is due to the muon halo background.
A search for peaks in the $m_{miss}$ distribution has been performed in the signal region, in steps of 1 MeV/$c^2$ . The Rolke-Lopez statistical method has been applied in each $m_{miss}$ bin to find the 90$\%$ confidence intervals for a poissonian process with gaussian background. No significance above 3 sigma has been found.
Figure~\ref{fig:na62-2007} (right) shows the expected and observed upper limits at $90\%~C.L.$ on BR($K^{\pm} \rightarrow \mu^{\pm} N$) in $10^{-5}$ units at each HN mass.
The upper limit on the squared matrix element magnitude $|U_{l4}|^2$ as a function of the HN mass hypothesis has been evaluated from those set for the BR according to equation (\ref{eq:BRequation}). In figure~\ref{fig:world} (left) the upper limit on $|U_{l4}|^2$ at $90\%$~ $C.L.$ as a function of the HN mass is compared with those set from the NA62 analysis of 2015 data and from other HN production searches in $K^+$\cite{E949}\cite{KEK} and $\pi^+$\cite{TRIUMF}\cite{PIENU} leptonic decays.

\subsubsection{NA62 2015 data}

NA62 has been taking data since 2015 with a completely renovated beam line and experimental apparatus \cite{NA62/Det} to measure BR($K^+ \rightarrow \pi^+ \nu \overline{\nu}$) with 10$\%$ accuracy. 
The SM prediction of this BR is of O($10^{-10}$) with very small uncertainty \cite{SM}.
The only measurement available so far is compatible with SM within a large error \cite{Kplus}, therefore
a precise measurement of the BR would be a fundamental test of the SM flavour sector.

In order to suppress background channels with BR up to 10 orders of magnitude higher than the signal, the NA62 experimental strategy requires a high kinematic rejection power, an effective photon and a muon reduction and excellent particle identification.
The experiment, approved for running until the year 2018, exploits a secondary positive charged hadron beam with 75 GeV/$c$ momentum and 1$\%$ r.m.s. spread, produced by SPS proton interactions with a beryllium target.
Beam kaons are tagged by a differential Cherenkov counter with nitrogen radiator. 
Beam particle momenta are measured by a silicon pixel detector (GTK) under commissioning in 2015 and not used in this analysis. 
A 75 m long fiducial decay volume in vacuum follows the last GTK station.
A spectrometer, made of four chambers with straw tubes in vacuum and a dipole magnet, measures track directions and momenta in the decay region.
 A system of calorimeters detects photons at different polar angles from the beam axis. 
Pions and muons are distinguished by a ring imaging Cherenkov detector filled with neon gas and by muon detectors.
The nominal instantaneous beam particle rate is 750 MHz, due to $\pi^+$ (70$\%$), protons (23$\%$) and $K^+$ (6$\%$). About 13$\%$ of the kaons decay in the fiducial volume, leading to about 5 MHz nominal $K^+$ decay rate. 
A muon flux (beam halo) with 3 MHz nominal rate in the detector acceptance is produced by kaon and pion decays upstream of the vacuum tank. 

\begin{figure}
\centering
\includegraphics[height=2.2in]{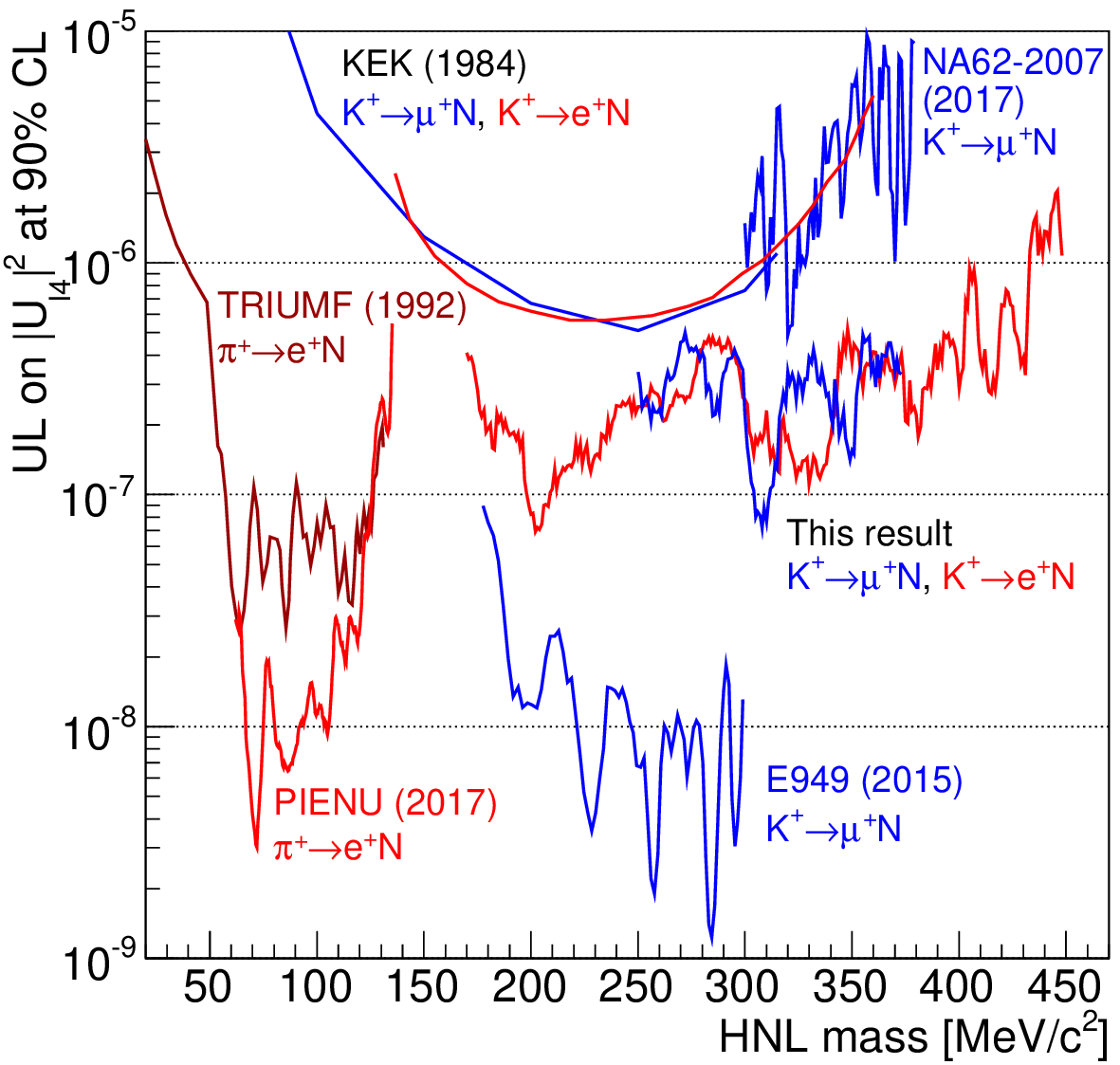}~~~~~~
\includegraphics[height=2.2in]{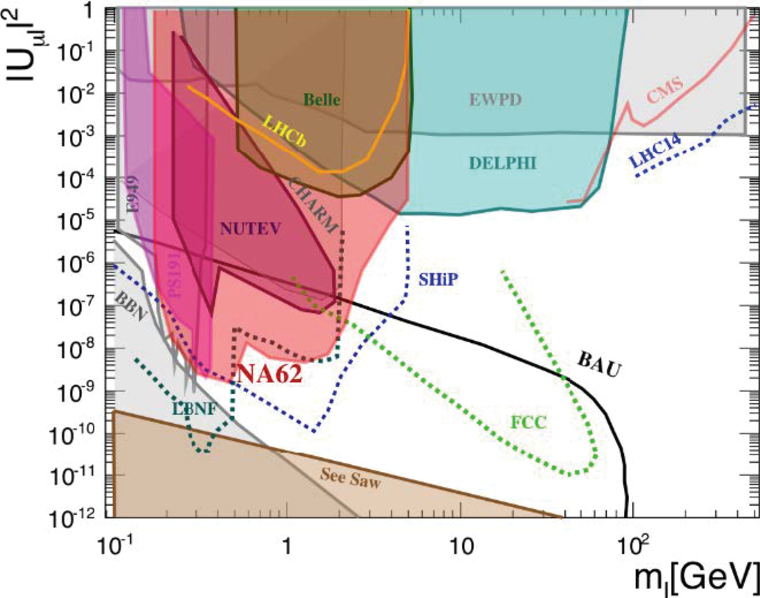}
\caption{Left: NA62 upper limits ($90\%$ ~$C.L.$) on $|U_{l4}|^2$ ($l = e, \mu$) as a function of the HN mass compared with previous limits established in HN production searches. Right: comparison of experimental and theoretical sensitivities (90$\%$ exclusion limit) for the HN-muon interaction strength $|U_{\mu}|^2$ as a function of HN mass.
}
\label{fig:world}
\end{figure}

Searches for HN production in $K^\pm\rightarrow l^\pm N$ ($l=e, \mu$) leptonic decays have been performed by NA62 with the data collected in a special minimum bias run in 2015 at 1$\%$ of the nominal beam intensity \cite{NA62/2015/HN}.  
About 300 millions kaon decays have been analyzed to search for HN in $K^\pm\rightarrow e^\pm N$ transitions in the mass signal region between 170 and 448 MeV/$c^2$. The search for HN in $K^\pm\rightarrow \mu^\pm N$ decays has been carried out in the mass signal region between 250 and 373 MeV/$c^2$ on a sample of about 100 million kaon decays.
No signal with significance above 3 sigma has been found.
New upper limits of $O(10^{-7})-O(10^{-6})$ have been established on the mixing matrix parameters $|U_{e4}|^2$ and $|U_{\mu 4}|^2$, improving the previous ones. 
Figure~\ref{fig:world} (left) shows the upper limits on $|U_{l4}|^2$ at $90\%$ ~$C.L.$ as a function of the HN mass hypothesis set by NA62 with the 2007 data and the final results of the 2015 data analysis on leptonic kaon decays, published shortly after the end of the NuPhys2017 conference, when only the analysis of the 2015 electronic decays was available and therefore presented. Results from HN productions searches in kaon\cite{E949}\cite{KEK} and pion\cite{TRIUMF}\cite{PIENU} leptonic decays are also quoted. 

Searches for very rare decays at NA62, as well as searches for new particles beyond the SM predictions, will benefit from the unprecedented size of the data sample, the excellent resolution on kinematic variables granted by the low material budget of the straw and GTK trackers, and the effective particle identification and photon veto capabilities of the NA62 detector. By the end of the run the limit on the mixing matrix parameters are expected to decrease by at least one order of magnitude.
The analysis of 2016 and 2017 data is well advanced and further improvements of the NA62 searches for HN are expected soon. 

\section{New projects for HN searches at the CERN SPS}
The SPS accelerator at CERN delivers to fixed target experiments the highest energy proton beam in the world, providing an integrated sample of $O(10^{19})$ particle-on-target (POT) per year.
The HNs predicted in the \nMSM~can be produced in proton interactions by secondary kaons, charmed and beauty mesons. 
Since hadron production cross sections increase steeply with energy,  CERN SPS is an excellent facility for experiments aiming at improving the current limits set by HN searches in meson decays up to few \GeVc2 mass values and down to $O(10^{-10})$ couplings. 
A detector placed as close as possible to a heavy material target and absorber, suited to reduce beam-induced background, and a decay volume tens of meters long, followed by a spectrometer with good particle identification capability, are main requirements for new experiments at the SPS aiming at investigating HN production and their subsequent decay. 
HN decays in SM particles have two charged particles in the final state.
Searches can be performed on 2-body decays or on 3-body open decays, with SM neutrinos in the final state. 
Since hadrons are totally absorbed by the dump material, large muon and neutrino fluxes from the decay of pions, kaons and short-lived resonances produced from proton interactions on the dump, are the main background sources.
The experimental challenge is to disentangle the signal from the huge muon and neutrino backgrounds that mimic its signature, spoiling the sensitivity to HNs.

The NA62 collaboration is currently considering the opportunity to run the experiment in 2022$-$2023, after the CERN accelerator shutdown foreseen in 2020$-$2021, to complete the  BR measurement and, afterwords, to take data in dump-mode collecting $10^{18}$ POT, in order to search for hidden particles from charm and beauty decays \cite{PBC0}\cite{PBC}\cite{NA62/Dump}.
Subject to approval, the SHiP (Search for Hidden Particles) collaboration proposes a new general purpose experiment to be installed in a dedicated beam line at the CERN SPS, aiming to collect $2 \times 10^{20}$ POT in 5 years of data taking starting from 2026 \cite{EPS2017-SHiP}\cite{ICHEP2016-SHiP}.
Both projects are part of the CERN Physics Beyond Collider (PBC) working group, providing inputs to the European Strategy for Particle Physics.

\subsection{NA62 in beam dump mode}
Thanks to the high intensity beam and the detector performance, NA62 can achieve sensitivities to light mediators in a variety of new-physics frameworks \cite{PBC0}\cite{PBC}\cite{NA62/Dump}.

Different scenarios are under test using data collected in the 2016 and 2017 run periods in different experimental conditions. 
Special low-bandwidth triggers have been implemented in parallel to the standard data taking. 
Short dedicated runs have been taken operating NA62 in dump-mode, with the beryllium target lifted and the proton beam dumped after closing the movable beam-defining Cu-Fe collimator installed 20 m downstream of the target to stop hadron interactions.
This operation is quick and fully reversible and can be easily exploited with the current setup.
In this configuration, with as little as $10^{18}$ POT, i.e. 80 days of data taking at the nominal beam intensity, about $2 \times 10^{15}$ D-mesons and $10^{11}$ b-hadrons can be produced.

The HN yield depends on the hierarchy of the active neutrino masses and on the relative interaction strength of the HN couplings to the three SM leptons $|U_{e}|^2$, $|U_{\mu}|^2$, $|U_{\tau}|^2$ \cite{HNTh}.
Figure~\ref{fig:world} shows, on the right, the expected NA62 sensitivity (pink shaded area) in dump-mode configuration with $10^{18}$ POT  for the coupling of HNs to muon neutrinos as a function of the HN mass for $|U_{e}|^2$ : $|U_{\mu}|^2$ : $|U_{\tau}|^2$ $\sim$ 1 : 16 : 3.8 and a normal hierarchy of active neutrino masses.
The allowed parameter space is limited by the BAU observation, the see-saw mechanism and the Big Bang Nucleosynthesis (BBN) lines; other shaded areas are excluded by past experiments, dotted lines are expected sensitivities of future experiments \cite{UmuHN}.
The NA62 sensitivity curve assumes to detect two-track final states from HN leptonic decays in $\pi e$ and $\mu e$ pairs with zero background and 100$\%$ selection efficiency, and includes geometrical acceptance and trigger efficiency.
Preliminary studies of background rates and topologies have been performed.
A rejection power down to zero background has been achieved at about $4 \times 10^{15}$ POT for fully reconstructed di-muon final states. 
Further studies are in progress to test background rejection at $10^{17}$ POT and with open decays.
Minimal improvements of the setup could be considered to optimize the detector design.

\subsection{The SHiP Project}
SHiP is a new general purpose fixed target experiment proposed to search for hidden particles at the CERN SPS and to complement the LHC experiments in the search for new physics \cite{EPS2017-SHiP}\cite{ICHEP2016-SHiP}.
Since hidden particles are expected to be accessible through heavy hadron decays, the facility is designed to maximize their production and the detector acceptance, while providing the cleanest possible environment. 

A dense target followed by a hadron absorber will stop all the SM particles but muons and neutrinos.
Further downstream, a challenging active muon shield is a critical component to strongly reduce the muon flux in the detector. 
A decay volume in vacuum, 60 m long, is foreseen, followed by a large magnetic spectrometer, calorimeters and particle identification detectors. Background taggers surrounding the decay vessel and a high resolution timing detector will improve the rejection of fake signal candidates. 
SHiP aims to probe different models predicting long-lived exotic particles with masses below $O(10)$ GeV/$c^2$ by integrating $ 2 \times 10^{20}$ POT in 5 years of data taking.
The SHiP sensitivity to HNs, compared in figure \ref{fig:world} (right) to experimental and theoretical sensitivities \cite{UmuHN}, is best up to 3 GeV, above the charm kinematic limit, thanks to a significant contribution from B-meson decays. In this region this sensitivity is unique and complementary to the region probed at the future circular collider (FCC) in $e^+e^-$ mode.
The CERN SPSC has reviewed the Technical Proposal and the Research Board has recommended SHiP to proceed with a Comprehensive Design Study to optimize the layout of the facility and the geometry of the detectors.

\end{document}